\documentclass[prb,twocolumn,showpacs,superscriptaddress]{revtex4}
\usepackage{bbm}
\usepackage{amsmath,amssymb}
\usepackage{graphicx}

\newcommand{\tr}{\mathop{\text{tr}}\nolimits}
\begin{document}

\title{Electron Bunching in Stacks of Coupled Quantum Dots}
\author{Rafael S\'anchez}
\affiliation{Instituto de Ciencia de Materiales, CSIC, Cantoblanco, 28049 Madrid, Spain}

\author{Sigmund Kohler}
\author{Peter H\"anggi}
\affiliation{Institut f\"ur Physik, Universit\"at Augsburg,
        Universit\"atsstra{\ss}e~1, 86135 Augsburg, Germany}

\author{Gloria Platero}
\affiliation{Instituto de Ciencia de Materiales, CSIC, Cantoblanco, 28049 Madrid, Spain}
\date{\today}

\begin{abstract}
We study the transport properties of two double quantum dots in a
parallel arrangement at temperatures of a few Kelvin.  Thereby, we
show that decoherence entailed by the substrate phonons affects the
shot noise.  For asymmetric coupling between the dots and the
respective lead, the current noise is sub-Poissonian for resonant
tunneling, but super-Poissonian in the vicinity of the resonances.
Our results indicate that the interaction between different channels
together with phonon emission and absorption are responsible for the
shot noise characteristics. The observed asymmetry of
the peaks at low temperatures stems from spontaneous emission.
\end{abstract}
\pacs{
73.63.Kv,  
73.23.Hk,  
72.70.+m,  
73.40.Gk   
}
\maketitle

\section{Introduction}
The aim of controlling and manipulating nanoscale devices requires
good knowledge of the processes involved in the electronic transport
through open quantum systems. The increasing success in accessing
single electron states in semiconductor quantum dots and the
unavoidable presence of lattice vibrations in such devices obliges one to
consider dissipation caused by electron-phonon interaction
\cite{leggett,tobias}. The study of the electronic current
fluctuations provides further information about the system
\cite{blanter,sigmund}.  E.g., from the investigation of shot
noise---a consequence of the charge discreteness---we know about
deviations from Poissonian statistics indicating correlations
between tunneling events.

A particular example for Poissonian statistics is the electron
transport through a point contact, for which all tunneling events are
statistically independent.  For resonant tunneling through single
quantum dots, this is no longer the case: As long as an electron
populates the quantum dot, no further electron can enter and,
consequently, tunneling events are anti-bunched.
However, when several of such transport
channels conduct in parallel and are coupled capacitively, the current noise becomes
super-Poissonian, as has been demonstrated experimentally
\cite{eth,zhang}.  This means that electrons tend to be transferred in
bunches, which at first sight is counterintuitive if one
thinks in terms of the Pauli exclusion principle. The phenomenon can
be understood in terms of Coulomb interactions between electrons in
different channels, so that an electron in one channel suppresses the
transport through the other \cite{cottet, gattobigio, yo},
and one observes \textit{dynamical channel blockade} (DCB).
Consequently the electron transport through one dot occurs in bunches
during lapses of time when the other dots are empty.

\begin{figure}[b]
\begin{center}
\includegraphics[scale=.6,clip]{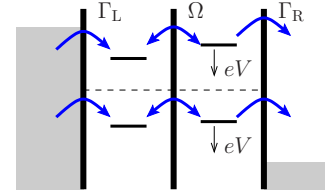}
\end{center}
\caption{\label{esquema} (color online)
Sketch of the transport through two parallel double quantum
dots measured in Ref.~\cite{barthold}.  Both transport channels are
capacitively coupled.  The source-drain voltage shifts the relative
position of the levels by $eV$, albeit it is so large that all levels
lie within the voltage window.}
\end{figure}
In a recent experiment \cite{barthold} with transport channels that
consist of double quantum dots (see Fig.~\ref{esquema}), intriguing
noise properties have been observed: By slightly modifying the
source-drain voltage, the levels of a double quantum dot can be tuned
across a resonance which yields a current peak at whose center, the noise
is sub-Poissonian.  In the vicinity of such resonances, by contrast,
the noise is super-Poissonian such that the Fano factor assumes values
up to 1.5.  This structure becomes washed out with increasing
temperature, indicating the suspension of DCB by the interaction
with substrate phonons.
In this work, we show that a model with a single transport channel
qualitatively reproduces this behavior.  For a quantitative agreement
with the experimentally observed Fano factor and temperature
dependence, however, the capacitive coupling to a second, almost
identical channel is found to be essential.

\section{Model}
We start out by modelling a single transport channel of the setup
sketched in Fig.~\ref{esquema}.  The double quantum dot coupled to
fermionic leads and substrate phonons is described by the Hamiltonian
\cite{AguadoBrandes,Kiesslich}
\begin{equation}
{H}
={H}_0+{H}_\text{leads}+{H}_T+H_\text{e-ph}+H_\text{ph},
\end{equation}
where
${H}_0
=\sum_{l=\rm L,R}\varepsilon_{l} n_{l} +U n_\text{L} n_\text{R}
-\Omega({c}_\text{L}^\dagger{c}_\text{R}
 +{c}_\text{R}^\dagger{c}_\text{L})/2$
describes the coherent dynamics inside the double dot and $n_l$
denotes the population of dot $l=\text{L},\text{R}$.  Henceforth we will
assume that the Coulomb repulsion $U$ is so strong that only
the zero-electron state $|0\rangle$ and the states with one electron
in the left or the right dot, $|\rm L\rangle$ and $|\rm R\rangle$,
play a role.  The leads and the phonons are described by
${H}_\text{leads}=\sum_{l,k}\varepsilon_{lk}
n_{lk}$ and $ H_\text{ph}=\sum_\nu
\hbar\omega_\nu a_\nu^\dagger a_\nu$, respectively, where $n_{lk}$ is
the electron number in state $k$ in lead $l$ and $a_\nu$ is the
annihilation operator of the $\nu$th phonon mode.  The interaction
with the double dot is given by the tunneling Hamiltonian
$ H_T=\sum_{l,k}(\gamma_{l}{d}_{lk}^\dagger{c}_{l}
+\text{h.c.})$ and the electron-phonon coupling $ H_\text{e-ph}= \sum_\nu
(n_L-n_R) \lambda_\nu (a_\nu^\dagger + a_\nu)$\cite{tobias}.
By tracing out the leads and the bath
within a Born-Markov approximation, we obtain for the reduced density
matrix the equation of motion
\begin{equation}\label{mastereq}
\dot\rho = {\cal L}\rho
= \left({\cal L}_0+{\cal L}_T+{\cal L}_\text{e-ph}\right)\rho  .
\end{equation}
Introducing for the density matrix the vector notation
$\rho=(\rho_{00},\rho_\mathrm{LL},\rho_\mathrm{LR},\rho_\mathrm{RL},\rho_\mathrm{RR})^T$,
the Liouvillian reads
\begin{equation}\label{L1}
{\cal L}
=\frac{1}{\hbar}\left(\begin{array}{ccccc}
-\Gamma_{\rm L} & 0 & 0 & 0 & \Gamma_{\rm R} \\
\Gamma_{\rm L} & 0 & -\frac{i}{2}{\Omega} & \frac{i}{2}{\Omega} & 0  \\
0 & -\frac{i}{2}\Omega{+}A_+ & i\delta{-}B & 0 & \frac{i}{2}\Omega{-}A_- \\
0 & \frac{i}{2}\Omega{+}A_+ & 0 & -i\delta{-}B & -\frac{i}{2}\Omega{-}A_- \\
0 & 0 & \frac{i}{2}{\Omega} & -\frac{i}{2}{\Omega} & -\Gamma_{\rm R}\\
\end{array}  \right),
\end{equation}
where the detuning $\delta=\varepsilon_{\rm R}-\varepsilon_{\rm L}-eV$
depends on the source-drain voltage (or on the gate voltages in
lateral quantum dots)
and $E^2=\delta^2+\Omega^2$.
For the phonons, we assume an Ohmic spectral density $J(\omega) =
\pi\sum_\nu \lambda_\nu^2 \delta(\omega-\omega_\nu) =
2\pi\alpha\omega$ \cite{leggett}, so that their influence is
determined by the coefficients
\begin{align}
A_\pm &=
   2\pi\alpha\Omega
   \pm2\pi\alpha\delta\Omega\Big(\frac{2 k_B T}{E^2}
   -\frac{1}{E}\coth\frac{E}{2 k_B T}\Big) , \\
B&=
   4\pi\alpha\Big(\frac{2 \delta^2 k_B T}{E^2}
   +\frac{\Omega^2}{E}\coth\frac{E}{2 k_B T}\Big)+\gamma,
\end{align}
where $\gamma=\Gamma_{\rm R}/2$ stems from the additional decoherence
associated with the tunneling to the leads.
In consistency with the experiment of Ref.~\cite{barthold}, we have
assumed that the voltage is so large that the Fermi level of the
left (right) lead is well above (below) the energy of the left
(right) dot level.  Therefore it is sufficient to consider only
unidirectional transport from the left lead to the right lead\cite{elattari} described by
the effective tunneling rates $\Gamma_{l}$ which are proportional to
$|\gamma_l|^2$.

Within the same approximation, one can derive for the current,
defined as the time derivative of the charge in the right lead, the
expression
$ 
I = e\tr_\text{sys}[({\cal J}_+-{\cal J}_-)\rho_0] ,
$ 
where $\rho_0$ denotes the stationary solution of the master equation
\eqref{mastereq} and $\mathcal{J}_\pm$ are the superoperators
describing the tunneling of an electron from the right dot to the
right lead and back, respectively.  For unidirectional transport, they
read ${\cal J}_-=0$ and $\mathcal{J}_+\rho = (\Gamma_{\rm R}/\hbar)
\rho_{\rm RR} |0\rangle\langle 0|$.

The noise will be characterized by the variance of the transported net
charge which at long times grows linear in time, $\langle\Delta
Q_{\rm R}^2\rangle = St$.  For its computation, we introduce the operator
$\tr_\text{leads+ph}(N_{\rm R}\rho_\text{total})$ \cite{Novotny2004a,
franz} which resembles the reduced density operator and obeys
\begin{equation}\label{noiseeq}
\dot\zeta(t)={\cal L}\zeta(t)+\left({\cal J}_+-{\cal J}_-\right)\rho(t).
\end{equation}
One can show \cite{franz} that $\zeta$ has a divergent component which
is proportional to $\rho_0$ and does not contribute to the
zero-frequency noise $S$.  Thus $S$ is fully determined by the
traceless part $\zeta_\perp=\zeta_0-\rho_0\tr\zeta$.
In terms of $\rho_0$ and $\zeta_\perp$, the zero-frequency noise reads
\cite{franz}
\begin{equation}
S
= e^2 \tr_\text{sys}[2({\cal J}_+-{\cal J}_-)\zeta_\perp
                     +({\cal J}_++{\cal J}_-) \rho_0 ].
\end{equation}
A proper dimensionless measure for the noise is the Fano factor
$F=S/eI$ which equals unity for a Poisson process, while a larger
value reflects electron bunching.

\section{Transport through a single channel}
%
\begin{figure}[bt]
\begin{center}
\includegraphics[width=\linewidth,clip]{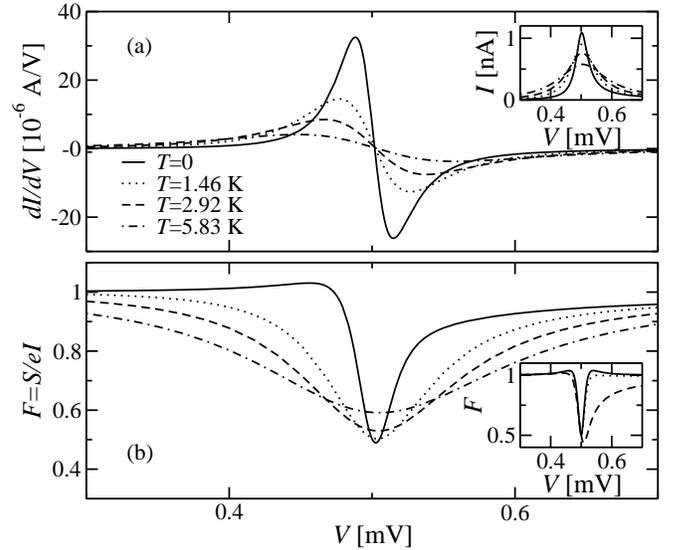}
\end{center}
\caption{\label{fig:onechannel}
(a) Differential conductance, current (inset), and (b) Fano factor
through a single channel for various temperatures and
$\Gamma_{\rm L}=0.025$, $\Gamma_{\rm R}=0.0125$, $\Omega=0.025$,
$\varepsilon=0.5$ and $\alpha=0.005$ (in meV). 
The inset of panel (b) shows the Fano factor for dissipation strength
$\alpha=0$ (solid), $10^{-3}$ (dotted), and $10^{-2}$ (dashed) at zero
temperature.
}
\end{figure}%
Figure~\ref{fig:onechannel}a shows the differential conductance and
the current for various temperatures as a function of the internal
bias.  In contrast to the dissipationless case ($\alpha=0$)
\cite{elattari,stoof}, the shape of the curve is no longer Lorentzian
but exhibits an asymmetry.  At higher temperatures, the peak
becomes broader and more symmetric.
This behavior is also reflected by the noise.  In the absence of
dissipation, the Fano factor deviates from the Poissonian value $F=1$:
For $\Gamma_{\rm L}>\Gamma_{\rm R}$ (as in the experiment)
and $\alpha=0$, we observe an antiresonant behavior with a dip
($F\approx 0.5$), which is accompanied by two maxima with values
slightly above 1. This double peak structure does not appear if
$\Gamma_{\rm L}\le\Gamma_{\rm R}$. With increasing dissipation
strength $\alpha$ and increasing temperature, the maxima
vanish and the Fano factor eventually tends to the Poissonian value
$F=1$.

Although this behavior resembles the experimental findings reported in
Ref.~\cite{barthold}, there are significant quantitative differences.
For the maximal peak value of the Fano factor, which is assumed in the
dissipationless limit $\alpha\to 0$ for $\delta=\Omega/\sqrt{2}$, we
find the analytic expression
\begin{equation}
F_p(\alpha=0)
=1+\frac{\Omega ^2 \left(\Gamma_{\rm L}-\Gamma_{\rm R}\right)^2}{2\Omega^2
(\Gamma_{\rm L} \Gamma _{\rm R}+2\Gamma_{\rm L}^2- \Gamma_{\rm R}^2) +8\Gamma_{\rm L}^2\Gamma_{\rm R}^2} .
\end{equation}
It implies $F_p \leq 5/4$, with the maximum assumed for $\Gamma_{\rm
R}\ll\Gamma_{\rm L},\Omega$.  This means that for a single channel,
the theoretical prediction for the maximal Fano factor is clearly
smaller than the value observed in the experiment even at finite
temperature and in the presence of dissipation \cite{barthold}; cf.\
inset in Fig.~\ref{fig:onechannel}b.  Therefore, we must conclude that
the one-channel model does not fully capture the experimentally
observed shot noise enhancement.

\section{Transport through two coupled channels}
The natural assumption is now that the shot noise must be influenced
also by the interaction with a second transport channel; cf.\
Fig.~\ref{esquema} and Ref.~\cite{barthold}.
Thus, we now consider two capacitively coupled channels, so that
the system Hamiltonian reads
$ H_0=\sum_{i,l} (\varepsilon_{il} n_{il}
+\frac{1}{2}\sum_{i',l'}U_{ii'll'} n_{il} n_{i'l'} )$,
where $i=1,2$ labels the different transport channels.  Note that
without interchannel interaction ($U_{ii'll'}=0$ for $i\neq i'$),
both channels are statistically independent.  Thus, the behavior
observed in the one-channel case (see Fig. \ref{fig:onechannel}) is repeated at a different voltage,
but still the Fano factor
cannot exceed the value $5/4$.

In order to simplify the model, we assume that the interaction
$U_{ii'll'}$ is huge whenever $i=i'$ or $l=l'$.  Then, the system will
accept up to two extra electrons provided that they are placed in
different stacks and different layers \cite{gattobigio,lambert}.  This
means that we have to consider the following seven states (the $i$th
letter refers to channel $i$): the empty state $|00\rangle$, the
one-electron states $|{\rm L}0\rangle$, $|{\rm R}0\rangle$, $|0{\rm
L}\rangle$, $|0{\rm R}\rangle$, and the two-electron states $|{\rm
RL}\rangle$, $|{\rm LR}\rangle$.
We assume that both dots on the right-hand side couple to the same
lead, while each channel couples to an individual phonon bath.  Then
we derive for the coupled channels a master equation of the form
\eqref{mastereq} with a Liouvillian given by an $11\times 11$ matrix.
A closer inspection of this Liouvillian reveals that---formally---it
can be obtained also in the following way: One writes the reduced
density operator of the double channel as a direct product of each
channel, $\rho = \rho^{(1)} \otimes \rho^{(2)}$, and the Liouvillian
accordingly as $\mathcal{L} = \mathcal{L}^{(1)} + \mathcal{L}^{(2)}$,
where $\mathcal{L}^{(i)}$ is the Liouvillian \eqref{L1} with the
parameters replaced by those of channel $i$. In this case,
$\gamma_i=(\Gamma_{j{\rm L}}+\Gamma_{i{\rm R}})/2$, $j\ne i$.
Finally, one removes all lines and columns that contain one of the
``forbidden'' states $|{\rm LL}\rangle$, $|{\rm RR}\rangle$.

For self-assembled quantum dots, a realistic assumption is that all
barriers are almost identical, so that $\Gamma_{{\rm L/R}}$ and
$\Omega$ do not depend on the channel index $i$.  By contrast, for the
internal bias $\varepsilon_i =\varepsilon_{i\rm R} -\varepsilon_{i\rm
L}$, we will find that already small differences play a role, so
that we have to maintain the channel index $i$ in the effective
detunings $\delta_{i} =\varepsilon_{i}-eV$.
For unidirectional transport, the current operators now read ${\cal
J}_-=0$, while $\mathcal{J}_+ = {\cal J}_+^{(1)} + {\cal J}_+^{(2)}$
acts on the reduced density operator as
$\mathcal{J}_+\rho = (\Gamma_{\rm R}/\hbar)[(\rho_{{\rm R}0}+\rho_{0{\rm R}})
|00\rangle\langle 00| + \rho_{\rm RL} |0{\rm L}\rangle\langle 0{\rm
L}| + \rho_{\rm LR} |{\rm L}0\rangle\langle {\rm L}0| ] $.

\begin{figure}[tb]
\begin{center}
\includegraphics[width=\linewidth,clip]{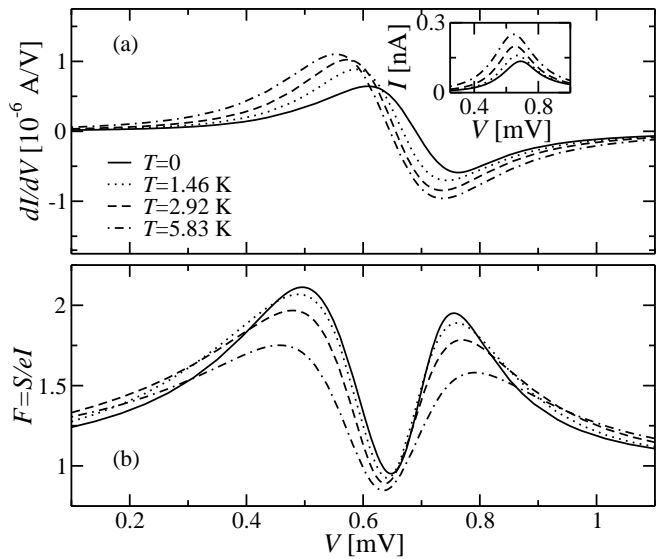}
\end{center}
\caption{\label{ddqd}
(a) Differential conductance and (b) Fano factor
for two coupled channels for various temperatures
and the parameters
$\Gamma_{\rm L}=0.025$, $\Gamma_{\rm R}=0.0125$, $\Omega=0.025$, $\alpha=0.005$,
$\varepsilon_{1}=0.5$ and
$\varepsilon_{2}=0.75$ (in meV).
The inset shows the temperature broadening of the current peak.
}
\end{figure}

In the absence of the phonons, we find the scenario discussed already
in Ref.~\cite{gattobigio}: The Fano factor exhibits two peaks, but
their origin is now different than in the one-channel case.  If both
double quantum dots become resonant at different source-drain
voltages, an electron in the double dot that is out of resonance has
only a small probability to tunnel through the central barrier.
Therefore, the non-resonant double dot will mostly be occupied with
one electron and thereby block the other channel, so that the current
peaks becomes smaller than in the one-channel case; cf.\ insets of
Figs.~\ref{fig:onechannel}a and \ref{ddqd}a.
Whenever the nonresonant channel is empty, however, the resonant
channel will transmit a bunch of electrons, so that eventually the
noise is super-Poissonian.

\begin{figure}[tb]
\begin{center}
\includegraphics[width=\linewidth,clip]{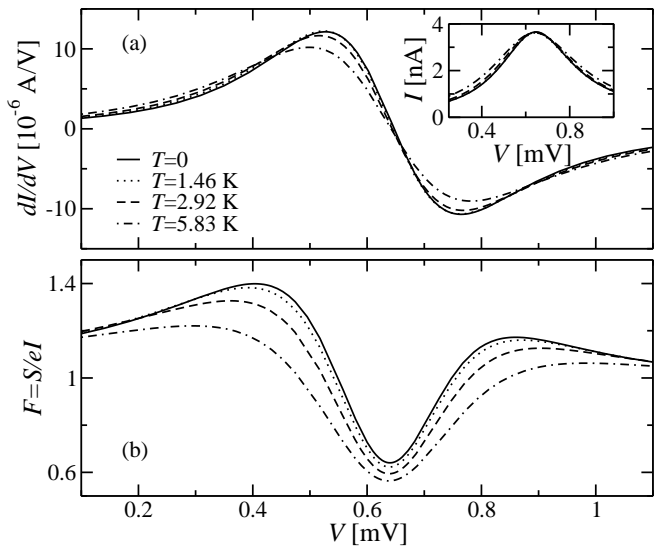}
\end{center}
\caption{\label{ddqd2}
(a) Differential conductance and (b) Fano factor
for two coupled channels for various temperatures
and the parameters
$\Gamma_{\rm L}=0.11$, $\Gamma_{\rm R}=0.055$, $\Omega=0.11$, $\alpha=0.005$,
$\varepsilon_{1}=0.5$ and
$\varepsilon_{2}=0.75$ (in meV).
The inset shows the temperature broadening of the current peak.
}
\end{figure}

Figures \ref{ddqd} and \ref{ddqd2} show the corresponding current and the Fano factor
in the presence of dissipation for two different configurations.  We observe two striking features
which are in accordance with the experimental results of
Ref.~\cite{barthold}: First, dynamical channel blocking is less
pronounced at higher temperatures and, second, the structure of the
Fano factor exhibits a clear asymmetry.

\begin{figure}[t]
\begin{center}
\includegraphics[scale=.5,clip]{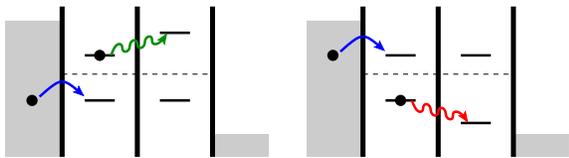}
\end{center}
\caption{\label{esqph} (color online) Phonon-assisted
channel opening: The blocking electron in the off-resonant channel can
tunnel through the interdot barrier after phonon absorption (left) or
emission (right) and thereby open the resonant channel.}
\end{figure}

This behavior can be explained within the following picture: Let us
consider, for instance, the situation sketched in Fig.~\ref{esqph}
where $\varepsilon_1<\varepsilon_2$.  When the source-drain voltage
puts the first double quantum dot in resonance, i.e., $\delta_1=0$,
the second double dot is still {\it above resonance} ($\delta_2>0$),
thus, blocking the resonant channel.  If now the electron in channel 2
absorbs a phonon, the blockade is lifted.  On the other hand, when
$\delta_2=0$, double dot 1 one is already {\it below resonance}
($\delta_1<0$) and phonon emission can resolve the blockade.
Both processes are more frequent the higher the temperature, so that
dynamical channel blockade is eventually resolved.  The fact that
emission is more likely than absorption, explains the observed
asymmetry and its reduction with increasing temperature.  We emphasize
that this does not rely on differences in the interdot hoppings
$\Omega_i$, in contrast to the mechanism of Ref.~\cite{gattobigio}.


This {\it phonon-induced channel opening} is also manifested in the
enhancement of the current shown in the inset of Fig.~\ref{ddqd}a.
The current peak becomes larger with increasing temperature and experiences a slight shift away in
its location. At low temperatures it tends to be around the larger resonance voltage ($\varepsilon_2/e$) which is driven by phonon relaxation. As phonon emission becomes important with temperature, the current peak becomes larger and shifts towards $eV_0=(\varepsilon_1+\varepsilon_2)/2$ coinciding with the maximal current voltage in the absence of dissipation.

This effect is weaker when considering stronger tunneling couplings: the Fano factor is reduced by DCB lifting and the current peak remains centered at $eV_0$ when increasing temperature, which only affects to its boadening, cf. Fig. \ref{ddqd2}. 

The observed behavior reproduces rather well the measurements reported
in Ref.~\cite{barthold}, but there is still one difference:
In the experiment, the Fano factor far from resonance is clearly
smaller than 1, while for the two-channel model, it tends to be
Poissonian.
This can be explained by leakage currents $I_k$ that inevitably
flow through the whole sample, but have been ignored so far.
We assume that the leakage currents are statistically
independent of each other and of the
coupled double dots considered.  Then we can write
both the current and the noise of the complete sample as a
sum of the independent channels:
$I_\text{sample}=I_\text{sys}+ \sum_k I_k$ and
$S_\text{sample}=S_\text{sys}+ \sum_k F_k I_k$,
where $F_k$ is the Fano factor associated to $I_k$.
If the leakage currents stem from resonant tunneling through single
quantum dots or double dots far from resonance, $F_i < 1$
(Ref. \cite{hershfield}) and, thus, the total Fano factor is decreased:
$F_\text{sample} =S_\text{sample}/I_\text{sample} <F_\text{sys}$.
However, since there are about $10^6$ leakage channels
\cite{barthold}, it is not possible to estimate their effect more
precisely.
   
\section{Conclusions}
To summarize, we have studied the effect of electron-phonon
interaction in the transport through double quantum dots systems,
predicting super-Poissonian shot noise whenever the source-drain
voltage tunes a double dot close to resonance.  The corresponding Fano
factor exhibits an asymmetric double-peak structure which becomes less
pronounced with increasing temperature.  In order to obtain the
experimentally observed values \cite{barthold} for the Fano factor, we
have assumed that two transport channels are so close that they can
block each other.  The temperature dependence of the double peaks have
been explained by the suspension of dynamical channel blocking via
phonon emission or absorption.  We attribute the sub-Poissonian noise
observed far from resonance to the appearance of independent leakage
currents.  So experiments with devices where these systems were
isolated from these additional noise sources are highly desirable.  

We thank F.~J. Kaiser for discussions. This work has been supported by
DFG through SFB 484 and the Spain-Germany Programme of Acciones
Integradas (DAAD and MEC).  S.K. and P.H. gratefully acknowledge support
by the German Excellence Initiative via the ``Nanosystems Initiative
Munich (NIM)''. R.S. and G.P. were supported by the M.E.C. of Spain
through Grant No. MAT2005-00644 and the UE network Grant No.
MRTN-CT-2003-504574.

\bibliographystyle{unsrt}

\end{document}